\documentclass[reprint,amssymb, amsmath,twocolumn, aps, superscriptaddress, showpacs, footinbib, pra, longbibliography	]{revtex4-2}
\usepackage{amssymb}
\usepackage{amsmath}
\usepackage{color}
\usepackage[colorlinks,bookmarks=false,citecolor=blue,linkcolor=blue,urlcolor=blue]{hyperref}
\usepackage{graphicx}
\usepackage{footmisc,booktabs}
\usepackage{gensymb}
\usepackage{sidecap}
\usepackage{microtype}
\usepackage{epstopdf}

\begin{document}

\title{Microscopic NMR evidence for successive antiferroelectric and antiferromagnetic order in the van der Waals magnet CuCrP$_2$S$_6$ }

\author{C. S. Saramgi}
\affiliation{Leibniz Institute for Solid State and Materials Research Dresden, 01069 Dresden, Germany}
\affiliation{Institute for Solid State and Materials Physics, TU Dresden, 01062 Dresden, Germany}
\author{L. F. Prager}
\affiliation{Leibniz Institute for Solid State and Materials Research Dresden, 01069 Dresden, Germany}
\author{S. Selter}
\affiliation{Leibniz Institute for Solid State and Materials Research Dresden, 01069 Dresden, Germany}
\author{Y. Shemerliuk}
\affiliation{Leibniz Institute for Solid State and Materials Research Dresden, 01069 Dresden, Germany}
\affiliation{Institute for Solid State and Materials Physics, TU Dresden, 01062 Dresden, Germany}
\author{S. Aswartham}
\affiliation{Leibniz Institute for Solid State and Materials Research Dresden, 01069 Dresden, Germany}
\affiliation{International Research Centre MagTop, Institute of Physics, Polish Academy of Sciences, al. Lotnik\'{o}w 32/46, 02-668 Warsaw, Poland}
\author{B. Büchner}
\affiliation{Leibniz Institute for Solid State and Materials Research Dresden, 01069 Dresden, Germany}
\affiliation{Institute for Solid State and Materials Physics, TU Dresden, 01062 Dresden, Germany}
\affiliation{Würzburg-Dresden Cluster of Excellence ct.qmat, Technische Universität Dresden, 01069 Dresden, Germany}
\author{H.-J. Grafe}
\email{H.grafe@ifw-dresden.de}
\affiliation{Leibniz Institute for Solid State and Materials Research Dresden, 01069 Dresden, Germany}
\author{K. M. Ranjith}
\email{ranjith.km1857@gmail.com}
\affiliation{Leibniz Institute for Solid State and Materials Research Dresden, 01069 Dresden, Germany}
\affiliation{Würzburg-Dresden Cluster of Excellence ct.qmat, Technische Universität Dresden, 01069 Dresden, Germany}
\affiliation{Center for Quantum Technologies and Complex Systems (CQTCS), Christ University, Bengaluru, Karnataka 560029, India}

\date{\today}

\begin{abstract}\noindent
We present a comprehensive $^{31}$P and $^{65}$Cu nuclear magnetic resonance (NMR) study of the layered van der Waals magnet CuCrP$_2$S$_6$. The compound exhibits a sequence of structural and magnetic phase transitions: a high-temperature paraelectric state, followed by a quasi-antiferroelectric (QAFE) state near 185 K, a long-range antiferroelectric (AFE) phase below 150 K, and finally, antiferromagnetic (AFM) order below $T_\mathrm{N}$ = 30 K. The evolution of the NMR spectra, NMR shift, and spin-lattice ($T_1^{-1}$) and spin-spin ($T_2^{-1}$) relaxation rates provide direct microscopic fingerprints of these transitions. The splitting of both the NMR line and $T_1^{-1}$ below the AFE transition demonstrates the emergence of two inequivalent P sites. From $K–\chi$ analysis, we extract nearly isotropic transferred hyperfine couplings and show that the NMR shift anisotropy originates primarily from the dipolar contribution, in contrast to Mn$_2$P$_2$S$_6$ and Ni$_2$P$_2$S$_6$. We determine the ferromagnetic intralayer exchange  $J_{\rm{intra}}\approx$ -4.9 K from the Curie–Weiss temperature, consistent with ferromagnetic layers antiferromagnetically stacked along the $c$-axis, and evaluate the Moriya high-temperature relaxation rate including cross-correlation effects of the P–-P dimer. Critical divergence of $T_1^{-1}$ near $T_\mathrm{N}$ yields a critical exponent $\gamma\simeq$ 0.45(4), placing CuCrP$_2$S$_6$ in a three-dimensional Heisenberg universality regime.
\end{abstract}

\maketitle
\section{Introduction}
\begin{figure*}[!htp]
  \centering
  \includegraphics[clip,width= 2\columnwidth]{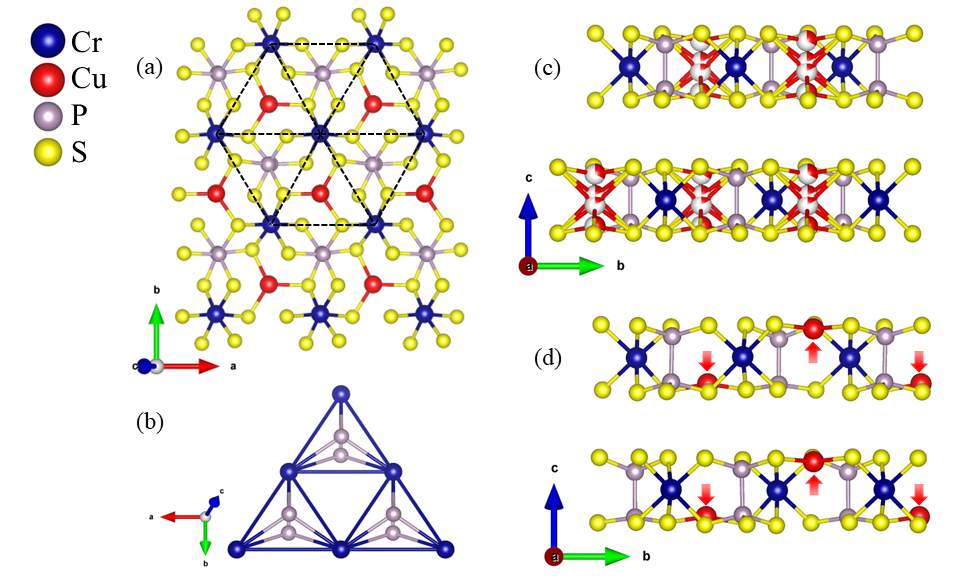}
  \caption{Crystal structure of CuCrP$_2$S$_6$. (a )View along the $c^{*}$ direction showing the CuCrP$_2$S$_6$ layer with Cr$^{3+}$ ions forming a two-dimensional triangular magnetic lattice within a network of [P$_2$S$_6$]$^{4-}$ units. (b): Local bonding environment of the P--P dimer. Each P–P dimer is coordinated to the same three Cr$^{3+}$ ions within the layer. (c) \& (d): Side view along the $a$ direction illustrating the stacking of quasi-2D layers separated by van der Waals gaps. (c): High-temperature ($T = 300$ K) configuration in the monoclinic $C2/c$ structure, where Cu ions occupy two (or four) partially occupied sites. (d): Low-temperature ($T = 64$ K) antiferroelectric structure, where Cu ions order into alternating up- and down-displacement positions, as indicated by the arrows, thereby lowering the crystal symmetry.}\label{structure}
\end{figure*}
Layered van der Waals (vdW) materials have become central to modern condensed-matter research due to their tunable physical properties and the possibility of isolating them down to the monolayer limit~\cite{novoselov20162,geim2013}. In particular, vdW magnetic materials provide a clean platform for exploring low-dimensional magnetism, anisotropic exchange interactions, spin--orbit coupling effects, and magnetoelectric coupling~\cite{huang2017,gibertini2019,gong2019,burch2018}. Even in their bulk form, the weak interlayer bonding preserves a pronounced quasi-two-dimensional (quasi-2D) character, enabling direct access to low-dimensional spin physics while still allowing the use of bulk-sensitive spectroscopic probes~\cite{xu2022}. These features have motivated intense research into vdW magnets for both fundamental studies and potential applications in spintronics and magnon-based information technologies~\cite{bonilla2018,zhao2025,zhang20242}.

Within this broader class, the transition-metal thiophosphates $T_2$P$_2$S$_6$ (T = Mn, Fe, Co, Ni, Cu, Cr) represent a particularly compelling family of quasi-2D vdW systems~\cite{brec1986,van2024}. Their structure consists of puckered layers of edge-sharing $T$S$_6$ octahedra connected by discrete [P$_2$S$_6$]$^{4-}$ ligand units, whose internal P--P dimers form a recurring structural motif~\cite{mayorga2017,cheng2021}. Strong intralayer interactions and weak interlayer van der Waals bonding give rise to diverse magnetic ground states depending on the transition-metal ion~\cite{synnatschke2019l,shemerliuk2021,Senyk2023,Abraham2023}. For example, Mn$_2$P$_2$S$_6$ exhibits N\'eel-type antiferromagnetic order with weak easy-axis anisotropy, whereas Ni$_2$P$_2$S$_6$ displays XXZ-type easy-plane antiferromagnetism and pronounced quasi-2D correlations~\cite{Wildes2015,Dioguardi2020,Bougamha2022,upreti2025}. These materials frequently show broad susceptibility maxima well above the N\'eel temperature, reflecting the onset of short-range antiferromagnetic correlations in the 2D limit. Their flexibility in chemical substitution and the presence of multiple ferroic degrees of freedom make the thiophosphates a versatile platform for exploring the interplay between structural, electric, and magnetic order.

Among this family, CuCrP$_2$S$_6$ is distinguished by the coexistence of electric-dipole and magnetic ordering within a narrow temperature range~\cite{selter2023,luo2025,hong2024}. At high temperature, CuCrP$_2$S$_6$ crystallizes in the monoclinic $C2/c$ space group, with nonmagnetic Cu$^{+}$ (3$d^{10}$) and magnetic Cr$^{3+}$ (3$d^{3}$, $S = 3/2$) ions alternating within each layer~\cite{luo2025,Susner2020}. The Cr$^{3+}$ ions form a two-dimensional triangular magnetic lattice in the layers [see Fig.~\ref{structure}(a)], while the Cu$^{+}$ ions occupy noncentrosymmetric positions within distorted sulfur octahedra. Structural studies report that, at room temperature, Cu$^+$ ions are distributed over multiple possible sites with partial occupancies [Fig.~\ref{structure}(c)]~\cite{Maisonneuve1993}. 


Upon cooling, the system undergoes a quasi-antiferroelectric (QAFE) transition near 185~K, followed by a long-range antiferroelectric (AFE) transition around 150~K~\cite{Maisonneuve1993,Maisonneuve1995,Cajipea1996,Studenyak2003,Susner2020,park2022,io2023,tang2025}. Neutron powder diffraction studies have established that below the AFE transition, CuCrP$_2$S$_6$ adopts a primitive monoclinic structure with space group $Pc$~\cite{Maisonneuve1993}, characterized by long-range antipolar ordering of Cu$^+$ ions occupying alternating up- and down-displaced positions along $\pm c^{*}$ within their distorted S$_6$ coordination environment [see Fig.~\ref{structure}(d)]. These Cu$^+$ displacements generate local ionic dipole moments whose antiparallel arrangement cancels the macroscopic polarization, thereby defining the AFE state. The Cr$^{3+}$ and [P$_2$S$_6$]$^{4-}$ framework undergoes only minor structural modifications across the transition, indicating that the dominant electric dipole contribution is primarily associated with the Cu$^+$ sublattice. 
Recent first-principles calculations have examined several candidate ordered configurations and identified a closely related antiferroelectric structure with space group $P2_1$ as the lowest-energy configuration~\cite{abraham2025}. Both structural descriptions capture the essential antipolar Cu ordering and associated symmetry lowering.

At $T_{\rm N} \approx 30$~K, the Cr$^{3+}$ sublattice undergoes antiferromagnetic ordering. Neutron and thermodynamic measurements establish that the ordered state consists of ferromagnetically aligned spins within each $ab$-plane that couple antiferromagnetically between planes, forming an A-type antiferromagnetic structure.\cite{Maisonneuve1993,Maisonneuve1995,Susner2020} Recent ESR studies demonstrate that strong short-range ferromagnetic correlations persist well above $T_{\rm N}$ and that the magnetic excitation spectrum hosts two distinct antiferromagnetic magnon gaps at zero field, reflecting the biaxial anisotropy inherent to CuCrP$_2$S$_6$~\cite{abraham2025}.

Despite extensive bulk and spectroscopic studies, a microscopic understanding of how the local magnetic and structural environments evolve across the QAFE, AFE, and AFM transitions remains incomplete. It is not yet clear how the symmetry breaking associated with Cu displacements and the development of magnetic correlations modify the local electronic and magnetic fields at the phosphorus sites. Nuclear magnetic resonance (NMR) provides direct access to these changes through the hyperfine interaction, which serves as a sensitive local probe of both structural distortions and spin dynamics. In related thiophosphates such as Mn$_2$P$_2$S$_6$ and Ni$_2$P$_2$S$_6$, $^{31}$P NMR has proven highly sensitive to local symmetry breaking, P-site inequivalence, and the development of quasi-2D magnetic correlations~\cite{Dioguardi2020,Bougamha2022}. Given that the P--P dimer is a central structural element of all $T_2$P$_2$S$_6$ compounds, and that CuCrP$_2$S$_6$ hosts multiple structural transitions that strongly affect the local phosphorus environment, a detailed NMR investigation provides a particularly powerful tool to probe the microscopic interplay between structure and magnetism.

In this work, we present a comprehensive $^{31}$P and $^{65}$Cu NMR study of single-crystalline CuCrP$_2$S$_6$ across its structural and magnetic transitions. Through temperature- and angle-dependent NMR spectra,  NMR shift analysis, $K$--$\chi$ correlations, and spin-lattice and spin-spin relaxation measurements, we resolve the signatures of the QAFE and AFE transitions. Our measurements reveal a clear splitting of the $^{31}$P spectra below $T_{\mathrm{AFE}}$, providing direct, site-resolved evidence for two inequivalent P sites arising from symmetry lowering in the antiferroelectric phase. This is accompanied by a corresponding splitting in $1/T_1$, demonstrating that the inequivalent sites probe the same magnetic fluctuations with different hyperfine couplings. In addition, the transverse relaxation $1/T_2$ exhibits a distinct evolution across the AFE transition, including the disappearance of the oscillatory dimer response, providing a dynamical signature of the freezing of the Cu dipole configuration. These results establish CuCrP$_2$S$_6$ as a rare example of a vdW material in which electric-dipole order and quasi-2D magnetism coexist and interact, and they place the compound within the broader context of the $T_2$P$_2$S$_6$ family of low-dimensional magnets.
We further present a quantitative analysis of the high-temperature relaxation within a Moriya-type framework that explicitly incorporates cross-correlations within the P--P dimer. This approach accounts for the observed factor-of-two enhancement of $1/T_1$ relative to the single-site estimate and provides a more general description of relaxation in systems containing strongly coupled nuclear dimers. Beyond the specific case of CuCrP$_2$S$_6$, our results also serve as a useful benchmark for the NMR community by illustrating how the NMR spectra, $1/T_1$, and $1/T_2$ evolve across an antiferroelectric transition and into a magnetically ordered state.

\section{EXPERIMENTAL}

NMR experiments using a standard spin-echo pulse sequence were performed on a well characterized CuCrP$_2$S$_6$ single crystal sample (\textbf{$3\times 0.6\times 0.1\times$ }mm$^3$), synthesized as described in Ref.~\cite{selter2023}.  $^{31}$P nucleus (nuclear spin $I=\frac{1}{2}$ and nuclear gyromagnetic ratio $\gamma_N$ = 17.236~MHz/T) provides a clean magnetic probe without quadrupolar complications. NMR spectra were obtained either by Fourier-transforming the echoes at fixed frequency or by summing echo intensities collected during frequency sweeps. The NMR shift, $K = [\nu-\nu_{ref}]/\nu_{ref}$ was calculated by taking the resonance frequency of the sample ($\nu$) with respect to the Larmor frequency of  $^{31}$P  nuclei ($\nu_{ref}$), calculated after the magnetic field calibration.

The nuclear spin–lattice relaxation curves were measured by an inversion recovery technique. Since $^{31}$P has the nuclear spin $I = \frac{1}{2}$, the value of $T_1$ at each temperature was estimated by fitting the recovery curve of the longitudinal magnetization to a single exponential function 
\begin{equation}
\label{spinhalf}
  \left(1-\frac{M(t)}{M(\infty)}\right)=A~\mathrm{exp}(-\frac{t}{T_1}).
\end{equation}
Here, $M(t)$ is the nuclear magnetization at a time t after the inversion pulse and $M(\infty)$ is the equilibrium magnetization.

We have also observed $^{65,63}$Cu signal at low temperatures, especially in the AFM ordered state. At high temperature, the $^{65,63}$Cu  signal is weak and hard to detect, probably due to fast relaxation. The low temperature recovery curve of the $^{65}$Cu longitudinal magnetization is fitted by the expression for the central ($+\frac{1}{2}\leftrightarrow -\frac{1}{2}$) transition of spin-$\frac{3}{2}$ nuclei~\cite{Abragam1961,Slichter1990,Narath1967},
\begin{equation}
\label{spin3half}
  \left(1-\frac{M(t)}{M(\infty)}\right)=0.1~\mathrm{exp}(-\frac{t}{T_1})+0.9~\mathrm{exp}(-\frac{6t}{T_1}).
\end{equation} 

Although the $^{65}$Cu nucleus is quadrupolar, the absence of resolved satellite transitions and the nearly single-exponential recovery of the central-transition signal allow the relaxation data to also be satisfactorily fitted using a single-exponential (Eq.~\ref{spinhalf}) function.

The spin–spin relaxation time $T_2$ was extracted from the decay of the transverse magnetization using a standard $\frac{\pi}{2} -\tau - \pi$ echo sequence. At high temperatures, the decay showed a characteristic oscillatory Gaussian–exponential form typical for P–P dimers, while in the AFE phase it evolves into a single exponential decay.

\section{RESULTS}
\subsection{NMR spectra}
\subsubsection{Paramagnetic state}
\begin{figure*}[!htp]
  \centering
  \includegraphics[clip,width=2\columnwidth]{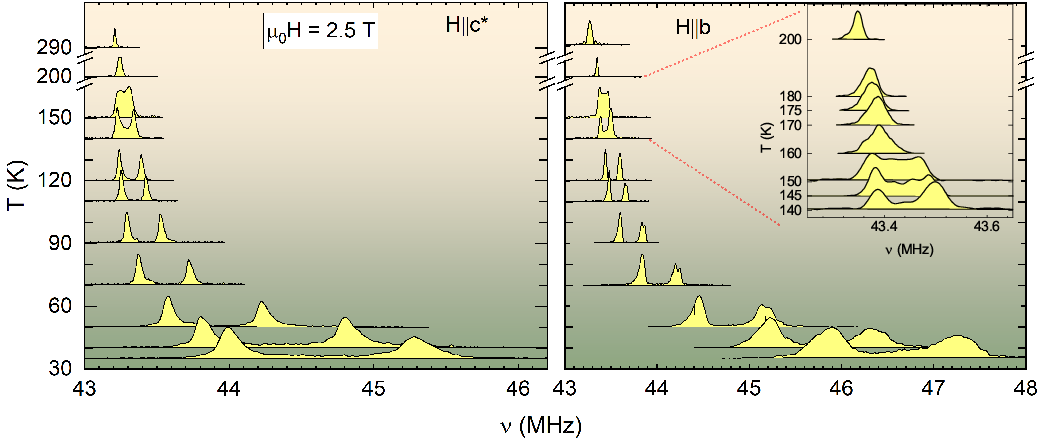}
  \caption{Temperature dependence of $^{31}$P NMR frequency spectra of CuCrP$_2$S$_6$ measured at 2.5~T applied along $c^*$-axis (left panel) and $b$-axis (right panel).   }\label{spectra}
\end{figure*}

The $^{31}$P nucleus ($I=\frac{1}{2}$) provides a clean magnetic probe without quadrupolar
complications, making the spectral line shape an excellent indicator of changes in
the local structural and magnetic environments.  
At high temperatures, where CuCrP$_2$S$_6$ crystallizes in the monoclinic
$C2/c$ structure, all P atoms are crystallographically equivalent and form
P--P dimers within the [P$_2$S$_6$]$^{4-}$ units.
Consistent with this structural simplicity, the $^{31}$P NMR spectrum consists of
a single, sharp resonance line without any detectable Pake doublet [see Fig.~\ref{spectra}]. A Pake doublet arises from the static homonuclear dipole–dipole interaction between two strongly coupled nuclear spins in a dimer, producing a characteristic two-peak splitting when the dipolar interaction is not dynamically averaged~\cite{Pake1948}.
The expected dipole--dipole splitting for the P--P dimer is on the order of a few kHz (approximately 3.5 kHz, as independently estimated from the spin-echo modulation discussed in Sec.~\ref{spinspin}), which is small compared to the magnetic linewidth arising from hyperfine coupling to the Cr moments in CuCrP$_2$S$_6$. As a result, the dipolar splitting is not resolved in the high-temperature phase. This interpretation is consistent with comparisons to related compounds. In Ni$_2$P$_2$S$_6$, where the linewidth is relatively small ($\sim$5~kHz), a clear Pake doublet is observed. In Mn$_2$P$_2$S$_6$, with a larger linewidth ($\sim$15~kHz), the Pake splitting is still visible but partially broadened~\cite{Dioguardi2020,Bougamha2022}. In contrast, in CuCrP$_2$S$_6$ the linewidth is significantly larger ($\sim$25~kHz), which exceeds the expected dipolar splitting and renders the Pake doublet unresolved.


\begin{figure}
  \centering
  \includegraphics[clip,width=1\columnwidth]{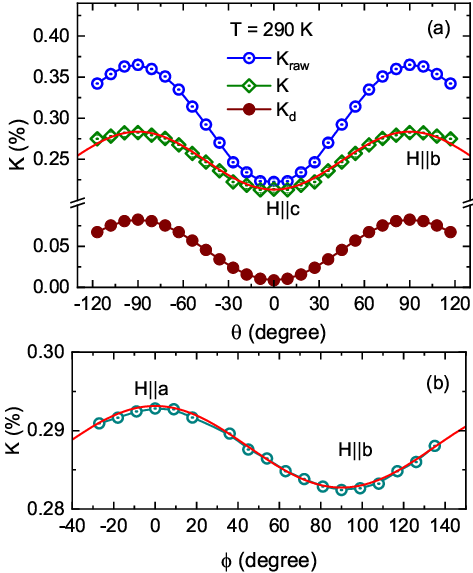}
  \caption{The measured $^{31}$P NMR shift $K_{\rm{raw}}$, contribution from bulk magnetic effects $K_{\rm{d}}$, and the intrinsic shift ($K = K_{\rm{raw}}-K_{\rm{d}}$) as a function of out-of-plane angle $\theta$, where $\theta$ = 0$^\circ$ represents ${\bf H}\parallel c^*$ and $\theta$ = 90$^\circ$ represents ${\bf H}\parallel b$ (b): K  as a function of in-plane angle $\phi$, where $\phi$ = 0$^\circ$ represents ${\bf H}\parallel a$ and $\theta$ = 90$^\circ$ represents ${\bf H}\parallel b$. Solid lines represent the fit using Eq.~\ref{Angular1} as described in the text.}\label{angular}
\end{figure}

 The angular dependence of the spectra at room temperature was investigated by performing in-plane ($a$ axis to $b$ axis, $\phi$) and out-of-plane ($c^*$ axis to $b$ axis, $\theta$) rotations.  To account for shifts caused by both bulk magnetization and demagnetization effects, we calculated the demagnetization contribution $K_d$ by assuming an ellipsoidal shape for the CuCrP$_2$S$_6$ crystal, as previously described~\cite{Dioguardi2020}. This contribution was then subtracted from the obtained NMR shift values. The resulting corrected NMR shifts $K$ are presented in Fig.~\ref{angular} for both out-of-plane ($\phi = 0$) and in-plane ($\theta$ = 0) rotations, where $\theta$ and $\phi$ are the polar and azimuthal angles, respectively. The angular dependence of the corrected NMR shift is well described by the general second-rank tensor expression,

\begin{equation}\label{Angular1}
\begin{split}
K(\theta,\phi) &= K_{\mathrm{iso}} 
+ \frac{K_{\mathrm{aniso}}}{2} (3\cos^{2}\theta - 1) \\
&\quad + \frac{K_{\mathrm{aniso}}\eta}{2} 
\sin^{2}\theta \cos 2\phi ,
\end{split}
\end{equation}

where $K_{\mathrm{iso}}=\frac{1}{3}(K_{a}+K_{b}+K_{c^*})$, is the isotropic shift, which is the orientation-independent average of the components.  $K_{\mathrm{aniso}}=K_{c^*}-K_{\mathrm{iso}}$ is the total shift anisotropy and $\eta=\frac{K_{a}-K_{b}}{K_{c^*}-K_{\mathrm{iso}}}$ is the asymmetry parameter that quantifies the deviation from uniaxial symmetry. $K_{a}, K_{b}$, and  $K_{c^*}$ are the three distinct principal components of the NMR shift tensor. The fit yields $K_{\mathrm{iso}}=0.263(4)~\%$,  $K_{\mathrm{aniso}}=-0.047(3)~\%$, and $\eta=0.22(1)$. The nonzero value of the $\eta$ reveals a moderate deviation from uniaxial symmetry, which differs from the behavior seen in other $T_2$P$_2$S$_6$ compounds~\cite{Dioguardi2020,Bougamha2022}. 

Figure~\ref{spectra} summarizes the evolution of the spectra for ${\bf H}\parallel b$ and ${\bf H}\parallel c^*$, at a fixed field of 2.5~T. Upon cooling from room temperature, the single high-temperature line undergoes a progressive shift reflecting the Curie--Weiss increase of the bulk spin susceptibility. As shown in the inset of Fig.~\ref{spectra}, below $\sim$180~K, the line begins to broaden significantly and develops an asymmetric shape. This regime corresponds to the onset of the quasi-antiferroelectric (QAFE) state, where Cu displacements become locally correlated but long-range order
has not yet been established. The broadening in this temperature range is attributed to a distribution of
local electric-dipole configurations, which modifies the local hyperfine fields at the P sites.
Such precursor fluctuations are also consistent with the damping of the oscillatory component of the spin--spin relaxation (see Sec.~\ref{spinspin}).

As the temperature is lowered further toward $T_{\mathrm{AFE}}\approx150$~K,
the broadened resonance resolves into two distinct peaks of approximately
equal intensity.
The emergence of two separate $^{31}$P lines directly signals a lowering of
the crystal symmetry and the formation of two inequivalent P sites in the
antiferroelectric (AFE) phase. 
The high-temperature $C2/c$ structure contains a single P site, the low-temperature $Pc$ phase exhibits four crystallographically distinct P positions. The two observed NMR lines therefore reflect the grouping of these sites into two distinct local hyperfine environments.

\begin{figure}
  \centering
  \includegraphics[clip,width=1\columnwidth]{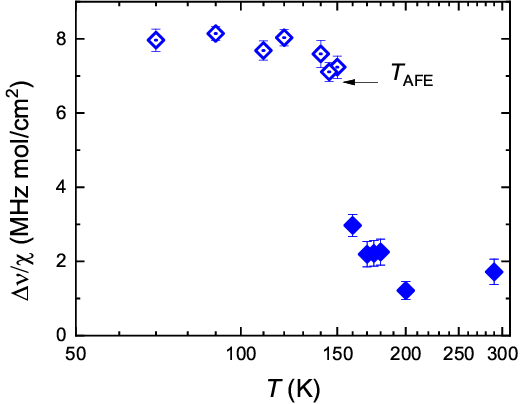}
  \caption{Temperature dependence of the $^{31}$P NMR linewidth and line splitting, both normalized by the bulk magnetic susceptibility, $\Delta \nu/\chi$, measured across the antiferroelectric (AFE) transitions. The linewidth is shown by filled symbols, while the line splitting is shown by open symbols.} \label{FWHM}
\end{figure}

The line splitting in the AFE state represents the order parameter of the AFE transition, the staggered polarization~\cite{Titze1998}. To disentangle magnetic and structural contributions to the NMR linewidth, we plot the temperature dependence of the  $^{31}$P  linewidth (line splitting) normalized by the magnetic susceptibility,   $\Delta\nu/\chi$ (Fig.~\ref{FWHM}). It is observed that the order parameter ($\Delta\nu/\chi$)  does not jump at 180~K, but instead grows continuously from zero at the transition temperature. Between 180 K and 150 K, the spectra exhibit an asymmetric line shape with continuous broadening and do not show signatures of phase coexistence between a fully developed AFE state with fixed splitting and a paraelectric component. In contrast, the clear line splitting observed below 150 K indicates the first-order nature of the AFE transition at 150 K. The presence of spectral intensity between the two split peaks in the temperature range 150–140 K further evidences the coexistence of fully AFE and paraelectric components and is another signature of a first order transition. This is consistent with the previous report on the powder sample~\cite{Moriya2005}.

The temperature evolution of the spectra therefore provides a microscopic
view of the structural transitions:  
(i) a single P site in the high-temperature paraelectric $C2/c$ phase,  
(ii) strong local distortions with local or short-range symmetry breaking in the QAFE regime (180–150 K), where Cu displacements remain short-range correlated before establishing long-range antiferroelectric order below 150~K, and  
(iii) two inequivalent P sites in the fully ordered AFE phase below 150~K. This sequence of transitions aligns well with calorimetric~\cite{Moriya2005} and structural measurements~\cite{selter2023}, demonstrating the high sensitivity of $^{31}\text{P}$ NMR spectroscopy to subtle symmetry changes in $\text{CuCrP}_2\text{S}_6$.

\subsubsection{NMR shift and hyperfine coupling}

\begin{figure}
  \centering
  \includegraphics[clip,width=1\columnwidth]{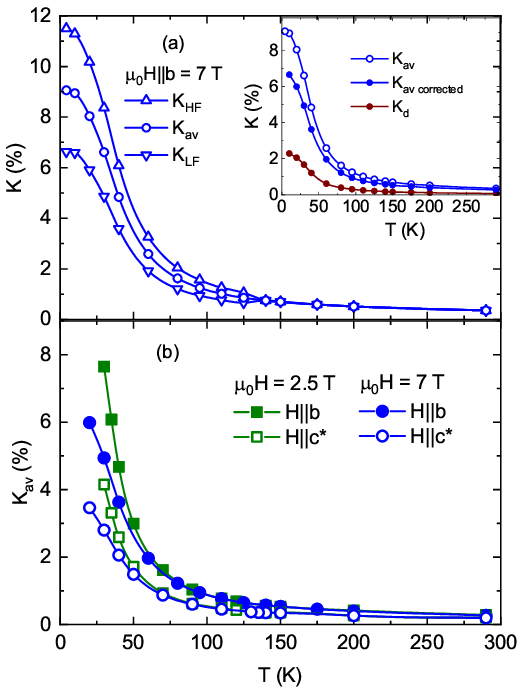}
  \caption{Temperature dependence of $^{31}$P NMR shift $K$. (a) NMR shift measured at 7~T applied along $c^*$-axis. $K_{\rm{HF}}$ and  $K_{\rm{LF}}$  are the shifts corresponding to high frequency and low frequency lines and   $K_{\mathrm{av}} = \frac{K_{\mathrm{HF}} + K_{\mathrm{LF}}}{2}$, represents the average shift. The inset shows the demagnetization correction of the shift, $K(T)= K_{\mathrm{av}}-K_{\mathrm{d}}$. (b): Corrected NMR shift $K(T)$  measured at 2.5~T and 7~T for ${\bf H}\parallel b$ and ${\bf H}\parallel c^*$. }\label{KvT}
\end{figure}

\begin{figure}
  \centering
  \includegraphics[clip,width=1\columnwidth]{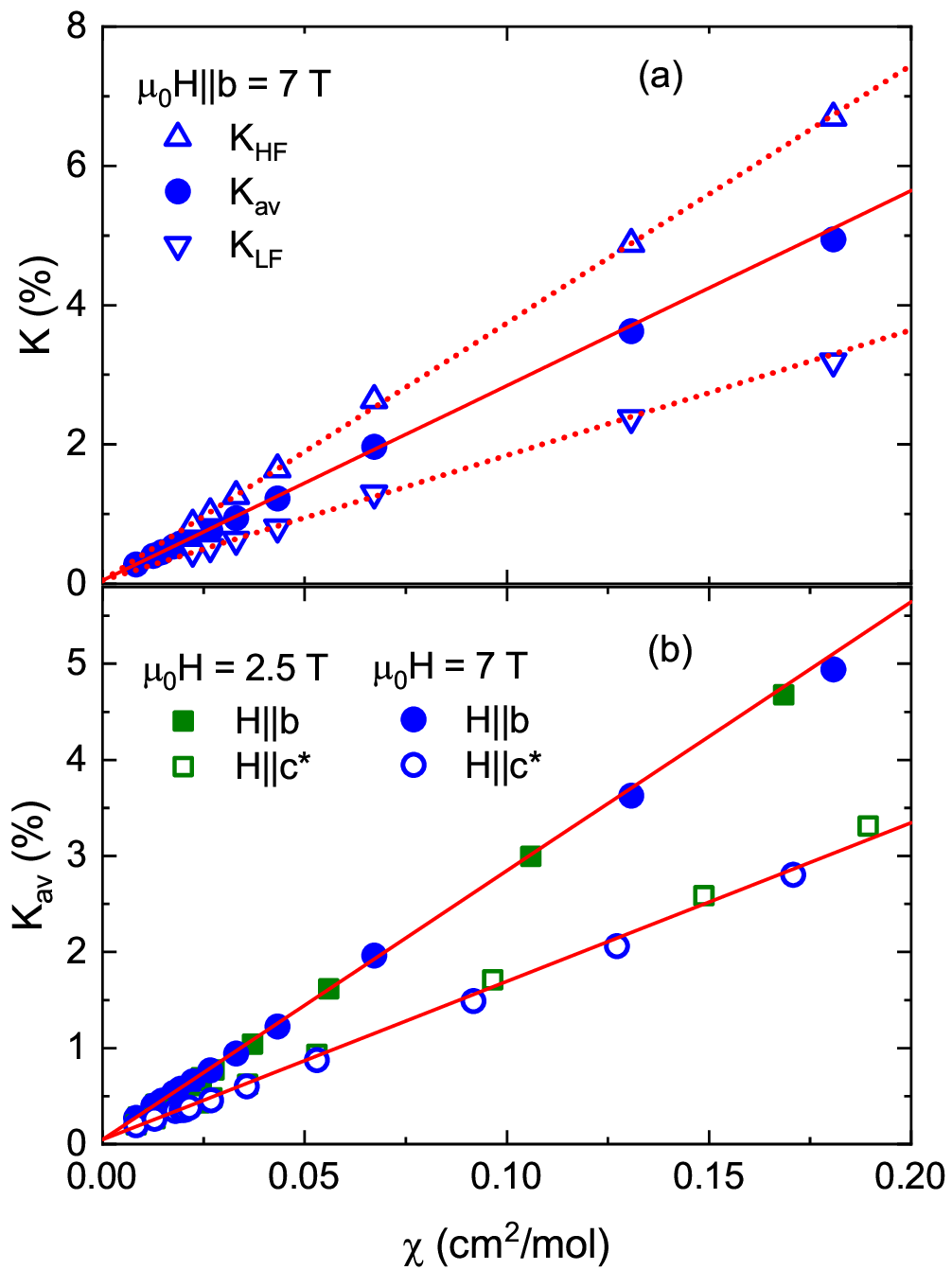}
  \caption{$^{31}$P NMR shift $K$ as a function of magnetic susceptibility $\chi$ with temperature as an implicit parameter. (a): $K \mathrm{vs}~ \chi$ plot at 7~T for the high frequency, low frequency, and average shifts. (b)  $K \mathrm{vs} \chi$ plot at 2.5~T and 7~T for ${\bf H}\parallel b$ and ${\bf H}\parallel c^*$. }\label{KvsChi}
\end{figure}

The NMR shift $K(T)$ provides a direct measure of the local spin
susceptibility at the P site and therefore serves as a sensitive probe of how
magnetic correlations evolve through the structural transitions.  
In the high-temperature paraelectric phase, the single crystallographic P
site yields a single NMR shift value that follows the bulk magnetic
susceptibility $\chi(T)$, indicative of a uniform hyperfine coupling between
the $^{31}$P nucleus and the fluctuating Cr$^{3+}$ moments.

Below the antiferroelectric transition at $T_{\mathrm{AFE}} \approx 150$~K,
however, the emergence of two inequivalent P sites leads to two distinct
NMR shift components, $K_{\mathrm{HF}}$ and $K_{\mathrm{LF}}$. As shown in Fig.~\ref{KvT}(a),
both exhibit a similar temperature dependence but differ in absolute magnitude,
reflecting differences in the transferred hyperfine fields arising from the
AFE distortion.
To characterize their common magnetic response, we define the average shift
\begin{equation}
K_{\mathrm{av}} = \frac{K_{\mathrm{HF}} + K_{\mathrm{LF}}}{2},
\end{equation}
which smooths out site-dependent offsets while preserving the intrinsic
temperature evolution. The intrinsic NMR shift $K(T)$ is obtained by subtracting the demagnetization correction $K_{\rm{d}}$ from the average shift $K_{\mathrm{av}}$ (inset of Fig.~\ref{KvT}(a)).  $K(T)$ measured at two different magnetic fields, 2.5~T and 7~T are shown in Fig.~\ref{KvT}(b), which follows a Curie--Weiss-like behavior over a wide temperature range, mirroring the bulk susceptibility.
This proportionality allows us to extract the transferred hyperfine coupling
constant $A_{\mathrm{hf}}$ using the Clogston--Jaccarino relation~\cite{Clogston1961}
\begin{equation}
K(T) = K_0 + \frac{A_{\mathrm{hf}}}{N_{\!A}\mu_B}\,\chi_{\mathrm{spin}}(T),
\label{eq:Kchi}
\end{equation}
where $K_0$ is the orbital (temperature-independent) shift and $\chi_{\mathrm{spin}}$
is the spin susceptibility obtained from bulk magnetization measurements. The transferred hyperfine interaction arises from spin density transferred through covalent bonding pathways between magnetic ions and ligand orbitals, producing an effective Fermi contact contribution at the ligand nucleus~\cite{Slichter1990,Abragam1961}.

As shown in Fig.~\ref{KvsChi}, plots of $K$ versus $\chi$ for both HF and LF sites are linear down to
$T_{\mathrm{AFE}}$, enabling reliable determination of $A_{\mathrm{hf}}$ and $K_0$. 
The parameters obtained from the fit using  Eq.~\ref{eq:Kchi} are tabulated in Table~\ref{Hyp}.
Strikingly, the orbital contribution $K_0$ is found to be essentially identical
for the two P sites below $T_{\mathrm{AFE}}$, indicating that the electronic
environment of the P atoms remains nearly unchanged through the structural
transition.
In contrast, the slopes of the $K$--$\chi$ plots, which directly yield
$A_{\mathrm{hf}}$, differ modestly between HF and LF sites.
This shows that the AFE distortion modifies the geometry of the Cr--S--P
superexchange pathways, thereby altering the transferred hyperfine coupling while leaving the local orbital environment nearly unchanged.~\cite{Slichter1990}.

\begin{table}
\caption{Hyperfine coupling constants $A_{\rm hf}$  and orbital shifts $K_0$ extracted from fits to $K$ versus $\chi$ for the field 7~T applied along $b$ and $c^*$ axes:  }
\begin{tabular}{cccc}
  \hline  \hline
&& $A_{\rm hf}$ (${T/\mu_{\rm B}}$)& $K_0$ (\%)\\  \hline 
&HF      &   0.22     &   0.045\\
(${\bf H}\parallel b$)&LF      &   0.1      &   0.045\\
&Av     &   0.16     &   0.045\\  \hline
&HF     &   0.13     &   0.03\\
(${\bf H}\parallel c^*$)&LF     &   0.046    &   0.036\\
&Av     &   0.1     &   0.048\\
  \hline
\end{tabular}
\label{Hyp}
\end{table}
\subsubsection{Magnetically ordered state}
\begin{figure*}
  \centering
  \includegraphics[clip,width=2\columnwidth]{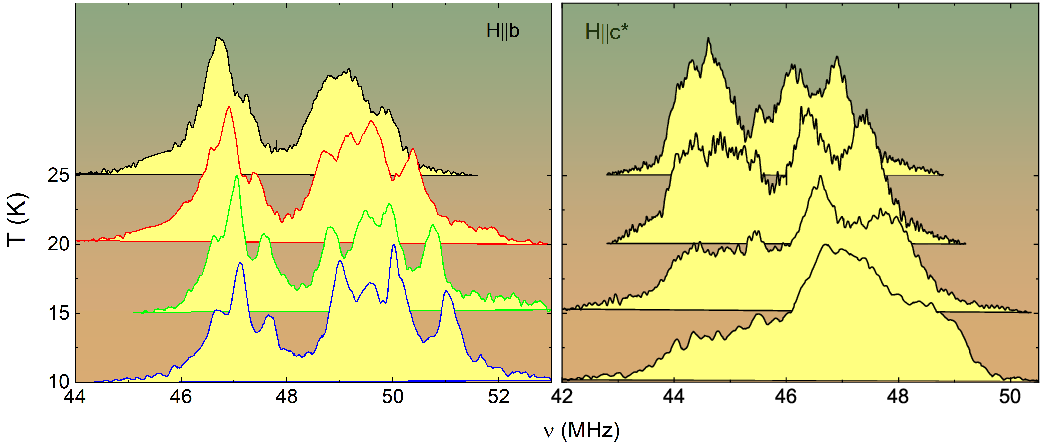}
  \caption{$^{31}$P NMR spectra in the antiferromagnetic state measured at 2.5~T for (a) ${\bf H}\parallel b$ and (b) ${\bf H}\parallel c^*$. }\label{spectra2}
\end{figure*}

At the Néel temperature \( T_\mathrm{N} = 30\, \mathrm{K} \), the Cr\(^{3+}\) sublattice in CuCrP\(_2\)S\(_6\) develops long-range antiferromagnetic order, which generates static internal hyperfine fields at the \(^{31}\)P sites. These fields produce additional splittings in the NMR spectra. Figure~\ref{spectra2} shows representative low-temperature spectra for magnetic fields along ${\bf H}\parallel b$ and ${\bf H}\parallel c^*$. Above \(T_\mathrm{N}\), the spectra show two peaks corresponding to the inequivalent P sites in the antiferroelectric phase. Upon cooling below \(T_\mathrm{N}\), each line splits due to static fields arising from ordered Cr\(^{3+}\) moments. For magnetic field along \(\mathbf{H} \parallel b\) (the magnetic easy axis), the splitting is  most pronounced, resulting in two pairs of lines for each phosphorus site. In contrast, \(\mathbf{H} \parallel c^*\) yields only a single pair of split lines. As in related thiophosphates such as MnPS$_3$ and NiPS$_3$, the additional lines in the antiferromagnetic ordered states are attributed to \(\pm 60^\circ\) stacking-related faults~\cite{Dioguardi2020,Bougamha2022}. In CuCrP$_2$S$_6$, however, the coexistence of the AFE splitting above $T_\mathrm N$ and the AFM splitting below $T_\mathrm N$ makes it difficult to unambiguously resolve the lines and quantify their relative populations reliably. We therefore restrict ourselves to the conclusion that stacking faults mainly affect the AFM internal-field pattern below $T_\mathrm N$, whereas the AFE splitting is intrinsic and already established above $T_\mathrm N$. A more detailed discussion of stacking-fault effects in related thiophosphates is given in Ref.~\cite{Dioguardi2020,Bougamha2022}.

\begin{figure}
  \includegraphics[clip,width=1\columnwidth]{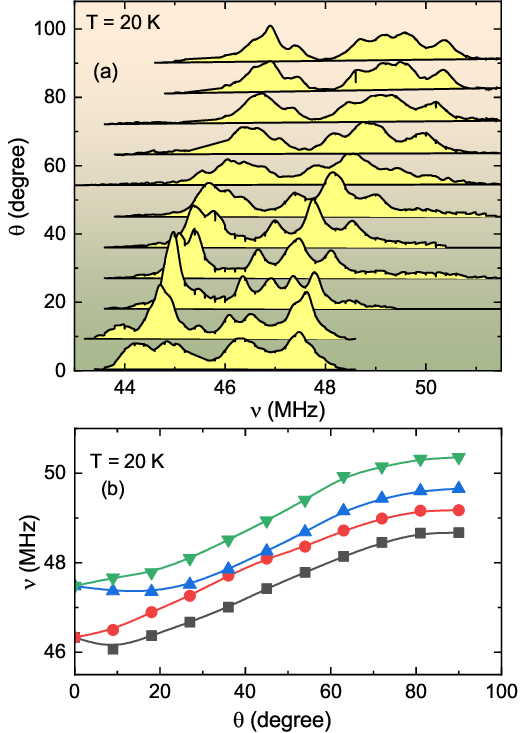}
  \caption{(a): Out-of-plane angular dependence of $^{31}$P NMR spectra in the antiferromagnetic state ($T$ = 20~K).  (b): Frequency vs $\theta$ extracted from multi-peak fit to the spectra in (a). }\label{spectra3}
\end{figure}

Angular-dependent NMR spectra recorded at 20 K [see Fig.~\ref{spectra3}(a)], as the crystal is rotated from the \(c^*\)-axis to the \(b\)-axis, were analyzed by multi-peak fitting. Extracted resonance positions exhibit clear angular dependence consistent with the anisotropic internal fields [Fig.~\ref{spectra3}(b)].

\subsection{Relaxation Measurements}
\subsubsection{Spin-lattice relaxation rate $T_1^{-1}$}

\begin{figure}
  \centering
  \includegraphics[clip,width=1\columnwidth]{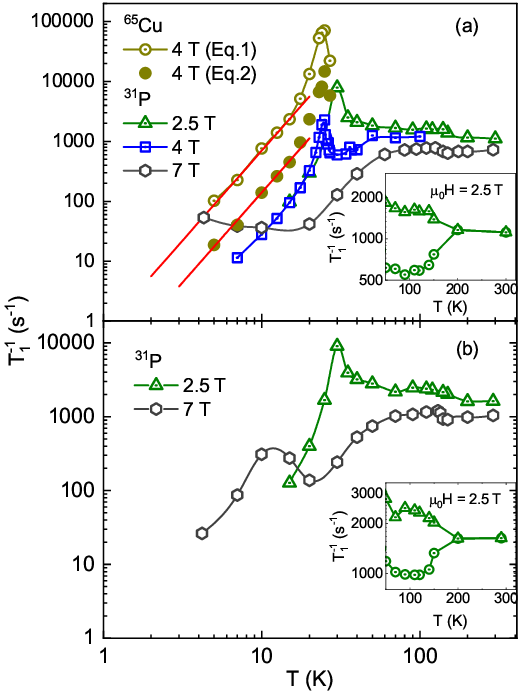}
  \caption{Temperature dependence of spin-lattice relaxation rate $T_1^{-1}(T)$ of CuCrP$_2$S$_6$ at different magnetic fields applied (a) ${\bf H}\parallel b$ and (b) ${\bf H}\parallel c^*$. The inset shows the splitting of  $T_1^{-1}(T)$  at $T_{\rm{AFE}}$. The solid line represents the  $T_1^{-1}(T)\propto T^3$ behavior.}\label{T1}
\end{figure}

Figure~\ref{T1}(a) and (b) show temperature dependence of the spin-lattice relaxation rate $T_1^{-1}(T)$ measured at different magnetic fields applied along $b$ and $c^*$ axes, respectively. At high temperatures ($T \gtrsim 180$~K), $T_1^{-1}(T)$ is nearly temperature independent, consistent with the exchange-narrowed limit expected for 
rapidly fluctuating Cr$^{3+}$ ($S=3/2$) spins in the paramagnetic state.  In this regime, the relaxation is dominated by the fluctuation spectrum of the local hyperfine fields arising from the three Cr neighbors within the 
same layer, and the nearly constant $T_1^{-1}$ reflects the high-frequency limit of the exchange field.

As seen in the inset of Fig.~\ref{T1}, a distinct splitting of the $T_1^{-1}(T)$ was observed at around 150~K, due to the different spin-lattice relaxation rates for each of the two split lines, providing clear microscopic evidence of the antiferroelectric phase transition. This phenomenon originates from the symmetry breaking that occurs as the system transitions from the high-temperature paraelectric phase to the low-temperature antiferroelectric phase. Above $T_{\rm AFE}$, in the more symmetric paraelectric state, all probed nuclei are crystallographically equivalent and thus exhibit a single, uniform relaxation rate. Upon cooling through the transition, the emergence of antiferroelectric order is accompanied by a reduction in crystal symmetry, which renders the formerly equivalent nuclear positions into two or more crystallographically distinct sites. Nuclei located in these new, inequivalent sites experience different local environments. Consequently, each distinct site is characterized by a unique spin-lattice relaxation rate due to variations in the local magnetic field fluctuations arising from different magnetic dipole-dipole interactions. The experimentally observed splitting of the $T_1^{-1}(T)$ is a direct manifestation of these multiple relaxation channels, confirming the lowering of symmetry at the transition. 
Furthermore, the distinct $1/T_1(T)$ values below $T_\mathrm{AFE}$ arise from their different transferred hyperfine couplings ($A_\text{hf}^\text{HF} \approx 0.22$~T/$\mu_{\mathrm B}$ vs.\ $A_\text{hf}^\text{LF} \approx 0.10$~T/$\mu_{\mathrm B}$ for $H \parallel b$). The measured ratio $(1/T_1)_\text{HF}/(1/T_1)_\text{LF} \approx 3-4.8$ (depending on exact $T$) closely matches $(A_\text{HF}/A_\text{LF})^2 \approx 4.8$, confirming that the symmetry breaking at 150~K produces multiple P environments with two different $A_\text{hf}$, but shared dynamical response.

Since $T_1^{-1}(T)$ for both LF and HF lines behaves similarly, but with different absolute values, we focus only on the $T_1^{-1}(T)$ of the HF line. Below $T_{\mathrm{AFE}}$, $T_1^{-1}$ remains nearly constant for both P sites 
over a broad temperature interval before increasing sharply upon approaching  the Néel temperature,  $T_{\mathrm{N}}$. A well-defined peak is observed at $T_{\mathrm{N}}=30$~K for both field orientations, signaling critical slowing down of the Cr-spin fluctuations.  We have also measured $T_1^{-1}(T)$ at various applied fields, and found that the position of the peak associated with the onset of magnetic ordering decreases with increasing magnetic field. The $T_1^{-1}(T)$ peak almost vanishes at $\mu_0H$ = 7~T for ${\bf H}\parallel b$, whereas the peak is visible at much lower temperatures ($\sim$10~K) for ${\bf H}\parallel c^*$. This is consistent with the thermodynamic measurements, where the magnetic saturation is reported as 6.5~T for ${\bf H}\parallel b$ and 7.8~T for ${\bf H}\parallel c^*$~\cite{abraham2025}.

At sufficiently low temperatures ($T \lesssim 25$~K), the $^{65}$Cu nuclei become observable, and the recovery of the longitudinal magnetization was analyzed using both single-exponential and bi-exponential relaxation models. While both fitting procedures yield comparable qualitative trends, $T_1^{-1}$  extracted from the single-exponential fits for $^{65}$Cu are more than one order of magnitude larger than those measured for $^{31}$P.
This is probably due to the strong hyperfine coupling between Cu nuclei and the Cr$^{3+}$ spins compared to that of P nuclei. Below $T_{\mathrm{N}}$,  $T_1^{-1}(T)$ decreases rapidly, following an approximate power-law behavior of $T_1^{-1}(T) \propto T^{3}$.   Such a $T^{3}$ dependence is characteristic of relaxation dominated by 
two-magnon Raman processes in a three-dimensional antiferromagnet with a gapped magnon spectrum~\cite{Belesi2006}. Similar power law dependence has been observed for several systems in the AFM-ordered state~\cite{Nath2014}. 

\subsubsection{Spin-spin relaxation rate $T_2^{-1}$}\label{spinspin}
 
\begin{figure}
  \centering
  \includegraphics[clip,width=1\columnwidth]{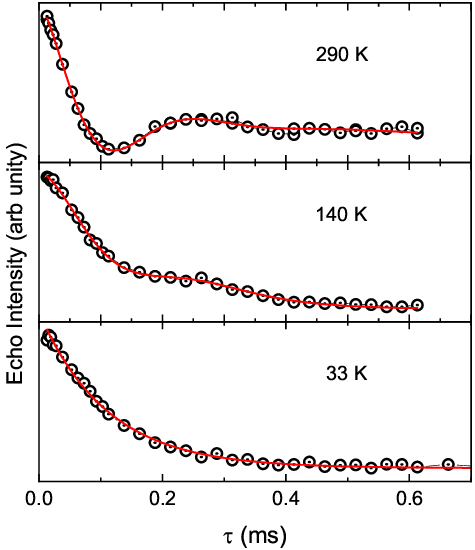}
  \caption{Transverse magnetization recovery curves at three selected temperatures along with the fits using Eq.~\ref{T2-osci} and Eq.~\ref{T2-expo}.}\label{T2-recover}
\end{figure}

  The transverse spin–spin relaxation rate $T_2^{-1}$ of $^{31}$P provides 
additional insight into the evolution of local magnetic and electric 
fluctuations across the structural and magnetic transitions.  
Representative spin–echo decays obtained at selected temperatures are shown in 
Fig.~\ref{T2-recover}.  
At high temperatures, the echo envelope exhibits pronounced oscillations, 
reflecting the dipolar coupling within the P–P dimer of the 
[P$_2$S$_6$]$^{4-}$ unit.  
The decay is well described by the standard Gaussian-modulated oscillatory 
form
\begin{equation}
\begin{split}
M(\tau)
&= M_0
\exp\!\left[-\frac{1}{2}\left(\frac{2\tau}{T_{2G}}\right)^2\right] \\
&\quad \times
\left[
1 - F\,e^{-2\tau/T_2}
\cos(2\omega_{\rm int}\tau - \psi)
\right],
\end{split}
\label{T2-osci}
\end{equation}
where $T_{2G}$ captures the static second moment of the local dipolar 
field distribution, $\omega_{\rm int}$ is the P–P dipolar oscillation frequency 
(\(\approx \gamma_n^2\hbar/r^3\)), and $F$ and $\psi$ parameterize the 
amplitude and phase of the oscillatory component.  
This functional form, which has also been reported in MnPS$_3$ and 
NiPS$_3$, reflects the strong homonuclear dipole-dipole coupling between the P nuclei in the
P--P dimers~\cite{Dioguardi2020,Bougamha2022}.

The internal field \(B_{\mathrm{int}}\) can be calculated from the fitted oscillation frequency using
\(B_{\mathrm{int}}=\omega_{\mathrm{int}}/{}^{31}\gamma\). In the high temperature para electric phase, the internal field is estimated to be  $B_{\mathrm{int}}\simeq$ 0.102~mT. 
This internal field can subsequently be used to estimate the expected homonuclear dipolar line splitting via:
\[
\Delta f=f_2-f_1
=2\left(\frac{{}^{31}\gamma}{2\pi}\right)B_{\mathrm{int}} .
\]
The resulting splitting is approximately \(3.5~\mathrm{kHz}\), which is significantly smaller than the observed linewidth and therefore remains unresolved in the spectrum.

Upon cooling below $\sim 180$~K, the oscillatory component becomes 
progressively damped, signaling the onset of 
slow fluctuations associated with the quasi-antiferroelectric (QAFE) 
regime.  
These fluctuations reduce the coherence of intra-dimer dipolar oscillations 
and represent the dynamical precursor to the formation of static 
Cu-displacement order.  
A more abrupt change appears at the long-range antiferroelectric (AFE) 
transition at $T_{\mathrm{AFE}}\approx150$~K, where the decay becomes a pure 
single exponential,
\begin{equation}
M(\tau) = M_0 \exp(-2\tau/T_2).
\label{T2-expo}
\end{equation}
The disappearance of the oscillatory component indicates that the local dipole configuration has frozen on the NMR time scale, consistent with the establishment of static AFE order.

\begin{figure}
  \centering
  \includegraphics[clip,width=1\columnwidth]{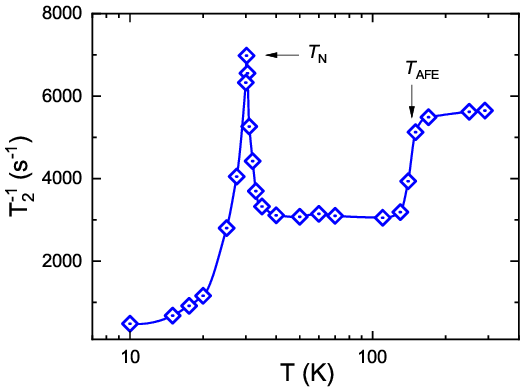}
  \caption{Temperature dependence of spin-spin relaxation rate $T_2^{-1}(T)$ measured at 2.5~T applied along ${\bf H}\parallel c^*$ . }\label{T2}
  \end{figure}

Figure~\ref{T2} illustrates the temperature dependence of $T_2^{-1}$measured at $\mu_0H$ = 2.5~T  applied along the $c$-axis. At high temperatures, $T_2^{-1}(T)$ remains nearly constant. Upon cooling, $T_2^{-1}$ drops below the AFE transition at ~150 K, without signatures of critical enhancement above this temperature.  This is also an indication of a first order transition.
 This decline halts at approximately 130~K, below which $T_2^{-1}(T)$ again becomes temperature-independent as the system settles into the antiferroelectric phase. With further decrease in the temperature, $T_2^{-1}(T)$ shows a sharp peak around 30~K, indicating the onset of antiferromagnetic long-range ordering. Below $T_{\mathrm{N}}$, $T_2^{-1}(T)$ decreases rapidly as transverse spin fluctuations freeze out.  
In the ordered state, the dominant relaxation mechanism becomes the dephasing induced by static field inhomogeneity and weak, residual magnon-mediated transverse fluctuations.  
 
\section{Discussion}

The NMR static and dynamic measurements provide compelling microscopic evidence for the sequential phase transitions in CuCrP$2$S$_6$, directly linking the spectral and relaxation rate changes to the emergence of antiferroelectric ordering. The narrow NMR line observed at room temperature is characteristic of a high-symmetry, paraelectric-like phase where $^{31}$P nuclei experience a single, dynamically averaged local environment. The onset of the quasi-antiferroelectric (QAFE) phase at 185 K is clearly marked by significant changes in the NMR spectra. The observed line broadening and developing asymmetry are indicative of a loss of local symmetry and the emergence of a distribution of static or slowly fluctuating electric dipole moments.
The transition into the long-range ordered antiferroelectric (AFE) phase below 150 K is unambiguously captured by the splitting of the NMR line into two distinct components. This spectral splitting is the definitive signature of a symmetry-breaking structural transition, confirming the formation of at least two crystallographically or electronically inequivalent nuclear sites in the AFE ground state. This result is fully consistent with the antiparallel displacement of ions that defines AFE order. The stabilization of this static order is further evidenced by the  nearly temperature-independent $T_2^{-1}(T)$ below 130 K, indicating that the fluctuations associated with AFE ordering have frozen out. 
Moreover, the observation of two distinct $T_1^{-1}(T)$ values for the two lines is attributed to different hyperfine coupling strengths at the corresponding sites, while the similar relaxation trends indicate that both sites experience the same underlying spin dynamics.

The hyperfine coupling between the phosphorus nuclei and the magnetic 3d ions in $T_2$P$_2$S$_6$ compounds such as Mn$_2$P$_2$S$_6$ and Ni$_2$P$_2$S$_6$ arises from comparable contributions of transferred and dipolar interactions, reflecting the relatively high density of magnetic ions within the lattice. To disentangle the hyperfine mechanism in CuCrP$_2$S$_6$, we calculated the dipolar hyperfine tensor $A_{\rm{dip}}$ at the P site using the high-temperature crystallographic structure (monoclinic $C2/c$), where all P sites are equivalent, and performed a lattice sum that extends up to 600~Å. The calculated dipolar hyperfine coupling tensor is given by

 \[
   A_{\rm dip}= 
  \left[ {\begin{array}{ccc}
   A_{xx} & A_{xy} & A_{xz} \\
   A_{yx} & A_{yy} & A_{yz} \\ 
   A_{zx} & A_{zy} & A_{zz} \\
  \end{array} } \right]
  =
    \left[ {\begin{array}{ccc}
   0.018 & 0.002 & 0.000 \\
   0.002 & 0.018 & 0.000 \\ 
   0.000 & 0.000 & -0.036 \\
  \end{array} } \right]
\]
where all values are given in units of T/$\mu_B$. These values are approximately a factor of two smaller than those in Mn$_2$P$_2$S$_6$ and Ni$_2$P$_2$S$_6$, reflecting the reduced number of magnetic ions in CuCrP$_2$S$_6$ (Cr$^{3+}$ occupies only half of the transition-metal sites compared to the Mn or Ni analogues). To determine the transferred hyperfine coupling, we subtract the calculated dipolar hyperfine couplings from the experimentally observed hyperfine values, i.e., $A_{\rm tra} = A_{\rm tot}-A_{\rm dip}$, where $A_{\rm tot}$ corresponds to the average value obtained from the $K$--$\chi$ analysis (Table~\ref{Hyp}).
The obtained transferred hyperfine coupling constants are $A_{\rm tra}^{\parallel b}=$0.142(4)~T/$\mu_{\mathrm B}$ and $A_{\rm tra}^{\parallel c^*}=$0.136(5)~T/$\mu_{\mathrm B}$ for ${\bf H}\parallel b$ and  ${\bf H}\parallel c^*$, respectively. It is interesting to note that the transferred hyperfine coupling constants are almost the same for ${\bf H}\parallel b$ and  ${\bf H}\parallel c^*$, indicating that the transferred contribution is isotropic. 
A comparison with related van der Waals magnets highlights the uniqueness of CuCrP$_2$S$_6$: in  Mn$_2$P$_2$S$_6$, the transferred hyperfine constants are much smaller (0.03 and -0.05 T/$\mu_{\rm B}$ for the $ab$ plane and $c$ axis, respectively) and only moderately anisotropic, while in Ni$_2$P$_2$S$_6$ the transferred coupling is both larger in magnitude and strongly anisotropic (-0.2 and -0.6 T/$\mu_{\rm B}$).

This isotropic nature of the $A_{\rm tra}$ in CuCrP$_2$S$_6$ strongly supports the prevalence of the Fermi contact interaction, which originates from spin density localized in the phosphorus $3s$ orbital and typically yields an isotropic coupling tensor. We attribute this selective spin-polarization mechanism to the electronic configuration of Cr$^{3+}$($3d^{3})$, where the unpaired magnetic electrons occupy the $t_{2g}$ orbitals and interact with the  [P$_2$S$_6$]$^{4-}$  framework primarily via $\pi$-bonding pathways. In contrast, in analogous compounds Mn$_2$P$_2$S$_6$ and Ni$_2$P$_2$S$_6$ unpaired electrons in the $e_g$ orbitals facilitate strong $\sigma$-bonding with the ligand framework, enabling direct spin transfer to anisotropic phosphorus 
$p$ orbitals and producing pronounced direction-dependent hyperfine interactions. Therefore, while the NMR shift anisotropy in Mn$_2$P$_2$S$_6$ and Ni$_2$P$_2$S$_6$  arises from a combination of anisotropic transferred and dipolar couplings, in CuCrP$_2$S$_6$ the anisotropy of the NMR shift originates exclusively from the dipolar contribution. This distinction highlights the different spin-polarization and bonding mechanism in CuCrP$_2$S$_6$ compared to other members of the $T_2$P$_2$S$_6$ family, which is manifested in the isotropic hyperfine coupling.

CuCrP$_2$S$_6$ exhibits a layered magnetic structure in which the Cr$^{3+}$ ($S=3/2$) spins are ferromagnetically aligned within each layer and antiferromagnetically coupled between adjacent layers. To estimate the exchange scales responsible for this ordering pattern, we extract the intralayer interaction from the bulk Curie--Weiss temperature. In the triangular Cr lattice, each spin interacts with six in-plane neighbors and two out-of-plane neighbors, leading to a Curie--Weiss temperature of $     \Theta_{\rm CW} = -\frac{S(S+1)}{3}\left( 6J_{\rm intra} + 2J_{\rm inter} \right)$.
Recent ESR measurements report an antiferromagnetic interlayer exchange of $J_{\rm inter} = +2.8~{\rm K}$~\cite{abraham2025}. Using the experimentally determined Curie--Weiss temperature $\Theta_{\rm CW} = 30~{\rm K}$~\cite{selter2023}, we obtain a ferromagnetic intralayer exchange of $J_{\rm intra} \approx -4.9~{\rm K}$. 
This negative value is consistent with the ferromagnetic alignment of spins within each Cr layer, while the positive interlayer coupling stabilizes the antiferromagnetic stacking of layers along the crystallographic $c$ axis.

In the high-temperature exchange narrowing limit of a Heisenberg magnet, the nuclear spin-lattice relaxation rate, $T_1^{-1}$, is determined by rapid exchange-mediated fluctuations of the local magnetic fields at the nuclear sites. In this regime, the relaxation rate becomes temperature-independent and can be expressed by the Moriya equation~\cite{Moriya1956}: 

\begin{equation}
\left(\frac{1}{T_1}\right)_{T\rightarrow\infty} =
\frac{(\gamma_{\mathrm{N}} g\mu_{\rm B})^{2}\sqrt{2\pi}z^\prime S(S+1)}{3\,\omega_{\mathrm{ex}}}
{\Big(\frac{A_{\perp}}{z'}\Big)^{2}},
\label{t1inf}
\end{equation}
where $S$  is the local moment spin, $\gamma_{\mathrm{N}}$ is the nuclear gyromagnetic ratio, $g$ is the electron $g$-factor, $A_{\perp}$ is the hyperfine coupling perpendicular to the applied field, $z^\prime$ is the number of nearest-neighbor Cr$^{3+}$ spins for a given P site, and $\omega_{\mathrm{ex}}$ is the Heisenberg exchange frequency, describes how fast the local fields fluctuate, set by the exchange motion among spins. The exchange frequency is defined as: 

$\omega_{\mathrm{ex}}=\left(|J_{\rm max}|k_{\rm B}/\hbar\right)\sqrt{2zS(S+1)/3}$, where $z$ is the number of nearest-neighbor spins of each Cr$^{3+}$ ion and $J_{\rm max}$ is the nearest-neighbor Heisenberg exchange constant. In CuCrP$_2$S$_6$, Cr$^{3+}$ form a triangular lattice arrangement in the $ab$ plane, where each Cr$^{3+}$  has six nearest-neighbours. Similarly, each P site is connected to three nearest-neighbour Cr$^{3+}$ spins. Thus using the parameters $\gamma_{N}$ = 108.303$\times$ 10$^2$ rad sec$^{-1}$, $A_{\rm hf}^{\perp}\approx$ 0.16~T/$\mu_{\rm B}$, $z$ = 6, $z^\prime$ = 3, $g$ = 2, $S=\frac{3}{2}$, $J_{\rm {intra}}\simeq$ 4.9~K, the high temperature value of spin lattice relaxation rate is calculated to be  $\left(\frac{1}{T_1}\right)_{T\rightarrow\infty}\simeq$ 503~sec$^{-1}$. On the other hand, the experimentally  observed $\frac{1}{T_1^{\parallel}}$ is  $\sim$1000~~sec$^{-1}$, which is almost a factor of 2 larger than the predicted value. Similar mismatches were also observed for other $T_2$P$_2$S$_6$ compounds~\cite{Bougamha2022}. 

In CuCrP$_2$S$_6$, the $^{31}$P nuclei form P--P dimers, and each P is coupled to the same three Cr$^{3+}$ spins with slightly different hyperfine tensors due to the local geometry. In Moriya's single-site theory, $T_1^{-1}$ is determined by the autocorrelation of the transverse local hyperfine field at a given nucleus, assuming statistically independent nuclear sites~\cite{Moriya1956}. In the present P--P dimer, however, the local fields at the two P sites originate from the same set of fluctuating Cr$^{3+}$ spins, so their temporal fluctuations are partially correlated~\cite{FreudeNotes}. Formally, the longitudinal relaxation of the two-spin system can be described by a $2\times 2$ relaxation matrix whose diagonal elements contain the usual single-site autocorrelation contributions, while the off-diagonal elements are proportional to the cross-correlation functions $\langle h_1^{+}(t) h_2^{-}(0)\rangle$ of the local fields at sites 1 and 2. For nearly equivalent sites, diagonalization of this matrix yields a fast symmetric mode with relaxation rate $\lambda_{+} = R_0 + R_c$ and a slow antisymmetric mode with $\lambda_{-} = R_0 - R_c$, where $R_0$ is the Moriya single-site rate and $R_c$ quantifies the cross-correlation. In the limit of strongly correlated fields ($R_c \approx R_0$), the symmetric mode relaxes at $\lambda_{+} \approx 2R_0$, whereas the antisymmetric mode is nearly protected, so that an NMR experiment probing the total unresolved $^{31}$P magnetization is dominated by the fast symmetric mode, leading to an effective $T_1^{-1}$ up to about twice the value calculated from the single-site Moriya expression, in agreement with our data~\cite{Ma2014}.

\begin{figure}
  \centering
  \includegraphics[clip,width=1\columnwidth]{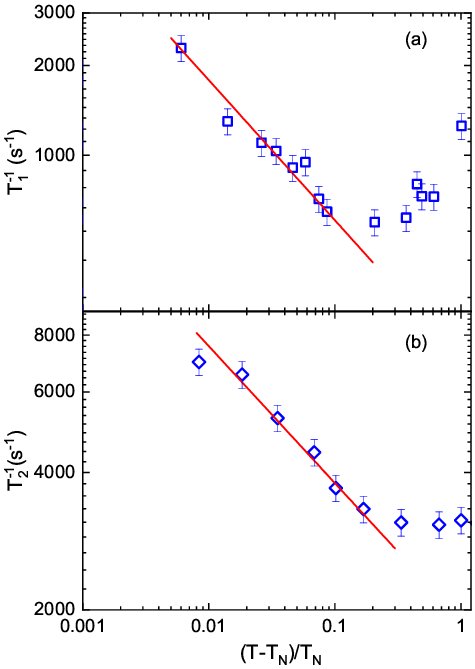}
  \caption{Log--log plots of the nuclear spin-lattice relaxation rate $T_1^{-1}$ measured at $\mu_0H = 4$ T and the spin-spin relaxation rate $T_2^{-1}$ measured at $\mu_0H = 2.5$ T as a function of the reduced temperature $(T-T_N)/T_N$. The corresponding field-dependent values of $T_N$ ($T_N = 24.9$ K for $T_1^{-1}$ and $T_N = 30.0$ K for $T_2^{-1}$) were used in calculating the reduced temperature.}\label{T1-critical}
\end{figure}

The nuclear spin lattice relaxation rate $T_1^{-1}(T)$  and spin-spin relaxation rate $T_2^{-1}$ show a critical divergence at the transition temperature. The singular behavior around the phase-transition point is characterized by the
power-law dependence of the physical quantities. In particular, $T_1^{-1}$ and $T_2^{-1}$ characterize the dynamical property of the phase transition. Figure~\ref{T1-critical} shows log-log plot of the nuclear spin lattice relaxation rate $T_1^{-1}$ and spin-spin relaxation rate $T_2^{-1}$ against the reduced temperature $\epsilon=(T-T_\mathrm{N})/T_\mathrm{N}$. It is well described by the power law, $T_{1,2}^{-1}\propto \epsilon^{-\gamma}$, for 0.01 $<\epsilon<$ 0.2, where $\gamma$ is the critical exponent. We obtained $\gamma=$ 0.45(4) for this fitting range. A mean field theory for a three dimensional isotropic Heisenberg antiferromagnet leads to $\gamma$ = 1/2~\cite{Moriya1962}. A dynamic scaling theory indicates $\gamma$ = 1/3 for a three dimensional isotropic Heisenberg model~\cite{Halperin1967}.  For comparison, larger exponents have been discussed for lower-dimensional systems, e.g. $\gamma \sim 0.8$ for a quasi-2D Heisenberg system~\cite{Takeya2008,deJongh2012} and $\gamma \sim 1.5$ for a 2D Ising system~\cite{Oyamada2008,collins1989}. The experimental value obtained here, $\gamma = 0.45(4)$, lies much closer to the range expected for 3D models than to those for 2D systems. Furthermore, the critical divergence is observed only within a relatively narrow reduced-temperature window, $0.01 < \epsilon < 0.2$, which is more characteristic of 3D criticality, whereas much broader critical regimes ($ \epsilon < 7$)are generally expected for 2D systems~\cite{itoh2009,Birgeneau1990}.

\section{Conclusion}
Our $^{31}$P and $^{65}$Cu NMR investigation provides a complete microscopic picture of the structural and magnetic transitions in the layered van der Waals magnet CuCrP$_2$S$_6$. The NMR spectra and relaxation rates reveal a sequence of ordering phenomena: a paraelectric state at high temperatures, a quasi-antiferroelectric regime near 185 K, long-range antiferroelectric order below 150 K, and antiferromagnetic order of Cr$^{3+}$ moments below 30 K.

NMR shift analysis demonstrates that the transferred hyperfine coupling is nearly isotropic, unlike in Mn$_2$P$_2$S$_6$ and Ni$_2$P$_2$S$_6$, and that the NMR shift anisotropy originates mainly from dipolar fields. The intralayer exchange interaction is determined to be ferromagnetic with $J_{\rm{intra}}\approx$ -4.9 K, consistent with ferromagnetic layers antiferromagnetically stacked along the $c$-axis.

The high-temperature deviation of $T_1^{-1}$  from Moriya’s single-site prediction is quantitatively explained by cross-correlation effects within the P--P dimer, leading to an enhancement of the symmetric relaxation channel. Near the magnetic transition, critical divergence of $T_1^{-1}$   with exponent $\gamma$= 0.45(4)  shows that CuCrP$_2$S$_6$ is consistent with the universality class of a three-dimensional Heisenberg antiferromagnet.

Together, these results firmly establish NMR as a powerful probe of coupled electric  and magnetic order in van der Waals materials and highlight the unique electronic and magnetic environment of CuCrP$_2$S$_6$ within the $T_2$P$_2$S$_6$ family.

\section*{Acknowledgment}
KMR acknowledges the financial support by the Deutsche Forschungsgemeinschaft
(DFG, German Research Foundation) under Germany’s Excellence Strategy through the Würzburg-Dresden
Cluster of Excellence on Complexity and Topology in Quantum Matter-ct.qmat (EXC 2147, project-id
390858490). SA acknowledges "MagTop" project (FENG.02.01-IP.05-0028/23) carried out within the "International Research Agendas" programme of the Foundation for Polish Science co-financed by the European Union under the European Funds for Smart Economy 2021-2027 (FENG). HJG acknowledges the financial support by the DFG through the Grant No: 534290830.
YS acknowledges financial support by the German Federal Ministry of Education and Research (BMBF) through project GU-QuMat (01DK240008).

\end{document}